\documentclass[%
 reprint,
superscriptaddress,
 amsmath,amssymb,
 aps,
]{revtex4-2}

\usepackage{amsmath}
\usepackage {xcolor}
\usepackage{comment}
\usepackage{bbm}
\usepackage{mathrsfs}
\usepackage{comment}
\usepackage{amssymb}
\usepackage{graphicx}
\usepackage{subfigure}
\usepackage[colorlinks,
            linkcolor=blue,
            anchorcolor=blue,
            citecolor=blue]{hyperref}

\newtheorem{corollary}{Corollary}
\newtheorem{proposition}{Proposition}

\allowdisplaybreaks[4]
\def\ra{\rangle}
\def\la{\langle}

\usepackage{graphicx}
\usepackage{dcolumn}
\usepackage{bm}

\begin{document}

\title{Quantum average correlations and complementarity relations via metric-adjusted skew information}

\author{Xiaoyu Ma}

\affiliation{College of Science, National University of Defense Technology, Changsha 410073, China}

\author{Qing-Hua Zhang}
\email[]{qhzhang@csust.edu.cn}
\affiliation{School of Mathematics and Statistics, Changsha University of Science and Technology, Changsha 410114, China}
\affiliation{Hunan Provincial Key Laboratory of Mathematical Modeling and Analysis in Engineering, Changsha University of Science and Technology, Changsha 410114, China}

\author{Cong Xu}
\affiliation{School of Mathematical Sciences, Capital Normal University,
Beijing 100048, China}


\begin{abstract}
We investigate quantum average correlations and complementarity relations based on metric-adjusted skew information. Several natural averaging procedures are considered, including complete families of mutually unbiased bases, all orthonormal bases, operator orthonormal bases, and twirling channels induced by the unitary group. All these approaches lead to the same closed expression, which identifies the resulting average correlation as an intrinsic quantity independent of the averaging scheme. By defining measures of wave and particle features via metric-adjusted skew information, we establish complementarity relations among wave and particle features, quantum entropy, and average correlation. These results provide a unified framework for investigating quantum average correlations and complementarity relations in terms of metric-adjusted skew information.
\end{abstract}

\maketitle



\section{Introduction}

Quantum correlations occupy a central position in quantum information science and are widely recognized as key resources in quantum computation, quantum communication, quantum cryptography, and quantum metrology \cite{Horodecki2009,Modi2012,Braun2018}. Entanglement provides the best-known paradigm of nonclassical correlations \cite{Horodecki2009}, while more general quantum correlations beyond entanglement can also yield operational advantages, for instance in mixed-state quantum computation and remote state preparation \cite{Datta2008,Dakic2012}. Accordingly, a variety of approaches have been developed to quantify quantum correlations, including entanglement measures \cite{Horodecki2009}, quantum discord and related entropy-based quantities \cite{Ollivier2001,Henderson2001,Modi2012}, as well as local-observable and incompatibility-based constructions \cite{PhysRevA.85.032117,Girolami2013}.

In the observable formulation, skew information quantifies the genuinely quantum part of the uncertainty of an observable and has been used as an operational measure of coherence \cite{Girolami2014Observable,Yu2017,LuoSun2017,Li2021,Sun2023}. More recently, increasing attention has been devoted to the interplay between quantum coherence and other nonclassical features, in particular quantum correlations and complementarity \cite{wwjf-lh44,fan2023average,fan2025average,Ma2016,PhysRevA.111.052451,PhysRevA.109.052439,PhysRevLett.116.160406,jin2021maximum,PhysRevLett.115.020403}. In particular, Luo and Sun further proposed average and maximal coherence based on the Wigner-Yanase skew information as quantifiers of quantum correlations \cite{LUO20192869,sun2017quantum}. At the same time, coherence has also been studied in close connection with mixedness and complementarity \cite{Zhang_2024}. This line of research was subsequently extended to the relative entropy of coherence and to other mixedness quantifiers, including those based on quantum entropies \cite{PhysRevA.92.042101,sun2022complementary,CHE2023106794,HU20181,PhysRevA.93.032136,PhysRevA.93.062111,PhysRevA.110.042413,wdxh-nwsw}.

The more general framework of metric-adjusted skew information was introduced by Hansen in connection with regular monotone Riemannian metrics~\cite{Hansen2008Metric,CaiHansen2010}. Within this framework, basis-independent coherence measures can be obtained by averaging over symmetric families, including mutually unbiased bases (MUBs), operator orthonormal bases, and all orthonormal bases \cite{fan2023average}. At the bipartite level, related ideas have also given rise to basis-free quantifiers of quantum correlations based on local observables, including formulations defined through summation over operator orthonormal bases, maximization over suitably normalized local observables, and minimization over local observables, as in local quantum uncertainty and its metric-adjusted extensions \cite{Gibilisco2021,fan2024quantification,sha2026quantifying,PhysRevA.110.022418,Sun2022QuantifyingCoherence}. These developments naturally motivate the study of average quantum correlations in the metric-adjusted skew information setting. 

In this work, we study average correlations and complementarity relations based on metric-adjusted skew information.The paper is organized as follows. In Sec.~\ref{sec2}, we introduce average coherence in the framework of metric-adjusted skew information. In Sec.~\ref{sec3}, we define quantum average correlation and establish the equivalence of its formulations based on mutually unbiased bases, operator orthonormal bases, all orthonormal bases, and twirling channels. In Sec.~\ref{wpd}, we quantify the wave feature and particle feature via metric-adjusted skew information. In Sec.~\ref{sec:complementarity}, we explore complementarity relations between wave-particle duality, quantum entropy and average correlation. Finally, Sec.~\ref{sec4} concludes the paper.

\section{Average coherence via metric-adjusted skew information}\label{sec2}

Let $\mathcal{H}$ be a $d$-dimensional Hilbert space. Denote by $L_\rho(X)=\rho X$ and $R_\rho(X)=X\rho$ the left and right multiplication operators, respectively. Recall Morozova-Chentsov function $c_f(L_\rho,R_\rho)$ associated with monotone Riemannian metrics on the quantum state space, the function $c_f(x,y)$ is given by \cite{Hansen2008Metric,GIBILISCO20092225}
\begin{equation*}
c_f(x,y)=\frac{1}{y f(x/y)}, \qquad x,y>0,
\label{eq2.2}
\end{equation*}
where $f:(0,\infty)\to(0,\infty)$ is a symmetric normalized operator monotone function satisfying
\begin{equation*}
f(1)=1,
\qquad
f(t)=t f(t^{-1}),
\qquad t>0.
\end{equation*}
If, in addition, $f(0):=\lim_{t\to 0^+}f(t)>0$, then the corresponding metric is regular.

For an operator $A$ and a state $\rho$ with spectral decomposition $\rho=\sum_{k=1} p_k |\psi_k\rangle\langle\psi_k|$, the metric-adjusted skew information associated with the Morozova-Chentsov function is defined by \cite{Hansen2008Metric,GIBILISCO20092225}
\begin{equation}\label{eq2.4}
\begin{aligned}
I_\rho^c(A)
&=
\frac{f(0)}{2}\,
{\rm Tr}\!\left(\mathbf{i}[\rho,A]\, c_f(L_\rho,R_\rho)\, \mathbf{i}[\rho,A]\right),
\end{aligned}
\end{equation}
where $\mathbf{i}=\sqrt{-1}$ is the imaginary unit.
Let
\[
{\tilde{c}_{f}}(x,y)=\frac{1}{2}\bigl[(x+y)-(x-y)^2 c_f(x,y)f(0)\bigr],
\]
then
\[
{\tilde{c}_{f}}(L_\rho,R_\rho)=\sum_{k,l}{\tilde{c}_{f}}(p_k,p_l)L_{|\psi_k\rangle\langle\psi_k|} R_{|\psi_l\rangle\langle\psi_l|}.
\]
Accordingly, the metric-adjusted skew information can be rewritten as \cite{FURUICHI20121147,10.1063/1.2748210,10.1090/S0002-9939-08-09447-1,GIBILISCO2011270,wdxh-nwsw}
\begin{equation}
\begin{aligned}
I_\rho^c(A)
&=
\frac{1}{2} {\rm Tr}\!\bigl[\rho(A^\dagger A + A A^\dagger)\bigr]
-
{\rm Tr}\!\bigl[A^\dagger {\tilde{c}_{f}}(L_\rho,R_\rho)A\bigr]\\
&=
\frac{1}{2} {\rm Tr}\!\bigl[\rho(A^\dagger A + A A^\dagger)\bigr]\\
&\quad -
\sum_{k,l}{\tilde{c}_{f}}(p_k,p_l)
{\rm Tr}\!\bigl[A^\dagger |\psi_k\rangle\langle\psi_k| A|\psi_l\rangle\langle\psi_l|\bigr].
\end{aligned}
\label{eq2.5}
\end{equation}
The quantity $I_\rho^c(A)$ is nonnegative and vanishes if and only if $[\rho,A]=0$. It is also convex in $\rho$ and thus provides a useful quantifier of quantum uncertainty and coherence.

Let
\[
\Pi=\{|i\rangle\langle i|: i=1,2,\ldots,d\}
\]
be a rank one von Neumann measurement associated with an orthonormal basis $\{|i\rangle\}_{i=1}^d$. The coherence of $\rho$ relative to $\Pi$ is then defined by \cite{Sun2022QuantifyingCoherence,fan2023average}
\begin{equation}
C^c(\rho\mid \Pi)
:=
\sum_{i=1}^d I_\rho^c(|i\rangle\langle i|).
\label{eq2.10}
\end{equation}
The quantity $C^c(\rho\mid\Pi)$ is nonnegative and vanishes exactly when $\rho$ is diagonal in the basis determined by $\Pi$. Since each term $I_\rho^c(|i\rangle\langle i|)$ is convex in $\rho$, the map $\rho\mapsto C^c(\rho\mid\Pi)$ is also convex.

To remove the explicit basis dependence, one may average Eq.~\eqref{eq2.10} over suitable families of measurements \cite{fan2023average}. A natural choice is a complete family of mutually unbiased bases. Recall that two orthonormal bases
$\{|b_{1j}\rangle: j=1,\ldots,d\}$,
and 
$\{|b_{2k}\rangle: k=1,\ldots,d\}$
are mutually unbiased if
\begin{equation*}
|\langle b_{1j}|b_{2k}\rangle|=\frac{1}{\sqrt d},
\qquad
1\le j,k\le d.
\end{equation*}
When $d$ is a prime power, there exists a complete set of $d+1$ mutually unbiased bases
\[ 
\Pi_t=\{|b_{tk}\rangle \la b_{tk}|: k=1,\ldots,d\},\ t=1,\ldots,d+1.
\]
The corresponding average coherence is defined by \cite{fan2023average}
\begin{equation}
C_{\rm MUB}^c(\rho)
:=
\frac{1}{d+1}\sum_{t=1}^{d+1} C^c(\rho\mid \Pi_t).
\label{eq2.13}
\end{equation}

An equivalent formulation can be given in terms of an operator orthonormal basis. Let $\mathcal{L}(\mathcal{H})$ denote the operator space on $\mathcal{H}$, and let $\{G_i\}_{i=1}^{d^2}\in \mathcal{L}(\mathcal{H})$ be an operator orthonormal basis satisfying
\[
{\rm Tr}(G_i G_j)=\delta_{ij}.
\]
The corresponding average coherence is defined by \cite{fan2023average}
\begin{equation}
C_{ob}^c(\rho)
:=
\frac{1}{d+1}\sum_{i=1}^{d^2} I_\rho^c(G_i).
\label{eq2.13b}
\end{equation}
This quantity is independent of the particular choice of the operator orthonormal basis and thus provides a basis free characterization of coherence at the operator level.

A further averaging procedure is obtained by averaging over all orthonormal bases, or equivalently over the unitary orbit of a fixed reference basis, that is,
\begin{equation}
C_{\mathcal U}^c(\rho)
:=
\int_{\mathcal U(d)} C^c\bigl(\rho\mid U\Pi U^\dagger\bigr)\, d\mu(U),
\label{eq2.14}
\end{equation}
where $d\mu(U)$ is the normalized Haar measure on the unitary group $\mathcal U(d)$ \cite{Sun2022QuantifyingCoherence}. By construction, $C_{\mathcal U}^c(\rho)$ is invariant under unitary conjugation of the state and depends only on the spectral properties of $\rho$ \cite{fan2023average,PhysRevA.105.032436}.

These three averaging procedures were shown to coincide \cite{fan2023average}. More precisely,
\begin{equation}
C_{\rm MUB}^c(\rho)=C_{\mathcal U}^c(\rho)=C_{ob}^c(\rho)
=
\frac{d-{\rm Tr}[{\tilde{c}_{f}}(L_\rho,R_\rho)]}{d+1}.
\label{eq2.15}
\end{equation}

Eq.~\eqref{eq2.15} implies that the resulting average coherence is independent of the choice of the complete family of mutually unbiased bases, and that it admits equivalent formulations in terms of an operator orthonormal basis and the Haar average over all orthonormal bases. Thus, averaging removes the explicit basis dependence of coherence while retaining nontrivial information about the state. This observation motivates the definition of quantum average correlation in bipartite systems.
\section{Quantum average correlation}\label{sec3}

We now extend the notion of average coherence to bipartite systems. Let $\rho^{AB}$ be a state on the composite Hilbert space $\mathcal{H}_A\otimes \mathcal{H}_B$, where $\dim \mathcal{H}_A=d_A$ and $\dim \mathcal{H}_B=d_B$. We write
\[
\rho^A={\rm Tr}_B(\rho^{AB})
\]
for the reduced state on subsystem $A$.

Let
\[
\Pi^A=\{|i^A\rangle\langle i^A|: i=1,2,\ldots,d_A\}
\]
be a rank one von Neumann measurement on $\mathcal{H}_A$. The corresponding local coherence of $\rho^{AB}$ with respect to $\Pi^A$ is defined by \cite{Sun2022QuantifyingCoherence}
\begin{equation}
C_A^c(\rho^{AB}\mid \Pi^A)
:=
\sum_{i=1}^{d_A} I_{\rho^{AB}}^c(|i^A\rangle\langle i^A|\otimes \mathbf{1}^{B}),
\label{eq3.1}
\end{equation}
where $\mathbf{1}^{B}$ is the identity operator on $\mathcal{H}_B$. This quantity measures the incompatibility between the bipartite state and the local measurement $\Pi^A$.

The corresponding basis dependent correlation is defined by \cite{PhysRevA.110.022418,Sun2022QuantifyingCoherence,PhysRevA.105.032436,sha2026quantifying}
\begin{equation}
Q^c(\rho^{AB}\mid \Pi^A)
:=
C_A^c(\rho^{AB}\mid \Pi^A)-C^c(\rho^A\mid \Pi^A),
\label{eq3.2}
\end{equation}
where
\[
C^c(\rho^A\mid \Pi^A)=\sum_{i=1}^{d_A} I_{\rho^A}^c(|i^A\rangle\langle i^A|)
\]
is the coherence of the reduced state $\rho^A$ with respect to the same measurement basis. Thus $Q^c(\rho^{AB}\mid \Pi^A)$ describes the part of the local coherence of $\rho^{AB}$ that cannot be accounted for by the marginal state $\rho^A$ alone.

Assume $\rho^{AB}$ has the spectral decomposition
\begin{equation}
\rho^{AB}=\sum_{k}\lambda_k |\Psi_k\rangle\langle\Psi_k|.
\label{eq3.8}
\end{equation}
For each $k$, $$|\Psi_k\rangle\langle\Psi_k|=\sum_{\mu, \nu}X_{\mu \nu}^k \otimes |\mu^k\ra \la \nu^k|,$$
where $X_{\mu \nu}^k:={\rm Tr}_B\left[|\Psi_k\rangle\langle\Psi_k| (\mathbf{1}^{A}\otimes |\nu^k\ra \la \mu^k|)\right]$. Denote by 
$$\sigma_k^B={\rm Tr}_A(\Psi_k\rangle\langle\Psi_k|)$$
the reduced state of $|\Psi_k\rangle\langle\Psi_k|$ on subsystem $B$.

\subsection{Average correlation over a complete family of MUBs}
Assume that $d_A$ is a prime power, so that there exists a complete family of $d_A+1$ mutually unbiased bases on $\mathcal{H}_A$. Let
\[
\Pi_t^A=\{|b_{ti}^A\rangle\langle b_{ti}^A|\!:\! i=1,2,\ldots,d_A\},
\ t=1,2,\ldots,d_A+1
\]
be such a complete family. The average correlation of $\rho^{AB}$ over this family is defined by
\begin{equation}
Q_{\rm MUB}^c(\rho^{AB})
:=
\frac{1}{d_A+1}\sum_{t=1}^{d_A+1} Q^c(\rho^{AB}\mid \Pi_t^A).
\label{eq3.4}
\end{equation}

\begin{proposition}\label{prop:mub}
For any bipartite state $\rho^{AB}$ on $\mathcal{H}_A\otimes \mathcal{H}_B$, the average correlation over a complete family of mutually unbiased bases is given by
\begin{equation}\label{prop2}
Q_{\rm MUB}^c(\rho^{AB})\!=\! \frac{{\rm Tr}[{\tilde{c}_{f}}(L_{\rho^A},R_{\rho^A})]\!-\!\sum_{k,l}{\tilde{c}_{f}}(\lambda_k,\lambda_l){\rm{Tr}}\left[ \sigma_k^B \sigma_l^B \right]}{d_A+1}.
\end{equation}
\end{proposition}

The proof of Proposition~\ref{prop:mub} is given in Appendix~\ref{app:mub}. Proposition~\ref{prop:mub} shows that $Q_{\rm MUB}^c(\rho^{AB})$ depends only on the state $\rho^{AB}$ and not on the particular choice of the complete family of mutually unbiased bases. Therefore, the MUBs average defines an intrinsic correlation quantity associated with the metric-adjusted skew information.

\subsection{Average correlation over all orthonormal bases}

We next average the basis-dependent correlation in Eq.~\eqref{eq3.2} over all orthonormal bases of subsystem $A$. Fix a reference rank-one von Neumann measurement
\[
\Pi^A=\{|i^A\rangle\langle i^A|:i=1,2,\ldots,d_A\}.
\]
Every orthonormal basis of $\mathcal H_A$ can be written as $U\Pi^A U^\dagger$ for some unitary operator $U\in \mathcal U(d_A)$. We therefore define
\begin{equation}
Q_{\mathcal U}^c(\rho^{AB})
:=
\int_{\mathcal U(d_A)}
Q^c(\rho^{AB}\mid U\Pi^A U^\dagger)\,d\mu(U),
\label{eq3.15}
\end{equation}
where $d\mu(U)$ is the normalized Haar measure on $\mathcal U(d_A)$.

By construction, the quantity in Eq.~\eqref{eq3.15} is independent of the chosen reference basis $\Pi^A$. One has the following closed expression for $Q_{\mathcal U}^c(\rho^{AB})$.

\begin{proposition}\label{prop:haar}
For any bipartite state $\rho^{AB}$ on $\mathcal{H}_A\otimes \mathcal{H}_B$, the average correlation over all orthonormal bases is given by
\begin{equation}
Q_{\mathcal U}^c(\rho^{AB})\!=\! \frac{{\rm Tr}[{\tilde{c}_{f}}(L_{\rho^A},R_{\rho^A})]\!-\!\sum_{k,l}{\tilde{c}_{f}}(\lambda_k,\lambda_l){\rm{Tr}}\left[ \sigma_k^B \sigma_l^B \right]}{d_A+1}.
\end{equation}
\end{proposition}

The proof of Proposition~\ref{prop:haar} is given in Appendix~\ref{app:haar}. Comparing the expressions in Propositions~\ref{prop:mub} and \ref{prop:haar}, we see that the average correlation over all orthonormal bases coincides with the average correlation over a complete family of mutually unbiased bases. Therefore, the average over all orthonormal bases yields an intrinsic correlation quantity depending only on $\rho^{AB}$ and the chosen metric-adjusted skew information.

\subsection{Equivalence of different average correlations}
Let $\mathcal{L}(\mathcal{H}_A)$ denote the real Hilbert space of Hermitian operators on $\mathcal{H}_A$, equipped with the Hilbert-Schmidt inner product. Let $\{G_i\}_{i=1}^{d_A^2}$ be an orthonormal basis of $\mathcal{L}(\mathcal{H}_A)$. The correlation based on an orthonormal basis is defined by \cite{PhysRevA.110.022418,Sun2022QuantifyingCoherence,sha2026quantifying}
\begin{equation*}
    \begin{aligned}
        Q_{ob}^c(\rho^{AB})&:=\frac{1}{d_A+1}\sum_{i=1}^{d_A^2} I_{\rho^{AB}}^c(G_i\otimes \mathbf{1}^{B}) - I_{\rho^A}^c(G_i)\\
&=\frac{{\rm Tr}[{\tilde{c}_{f}}(L_{\rho^A},R_{\rho^A})]\!-\!\sum_{k,l}{\tilde{c}_{f}}(\lambda_k,\lambda_l){\rm{Tr}}\left[ \sigma_k^B \sigma_l^B \right]}{d_A+1}.
    \end{aligned}
\end{equation*}

Specially, $\{\frac{\mathbf{1}^A}{\sqrt{d_A+1}},\frac{G_i}{\sqrt{d_A+1}}: i=1,2,\cdots,d_A^2\}$ is a set of Kraus operators of the depolarizing channel $\mathcal{E}_{\mathrm{De}}$. The metric-adjusted skew information for the depolarizing channel $\mathcal{E}_{\mathrm{De}}$ is defined by \cite{PhysRevA.105.032436,fan2023average,fan2024quantification}:
$$
C^c(\rho^{A} \mid\mathcal{E}_{\mathrm{De}}):=\sum_{i=1}^{d_A^2} I^c_{\rho^A}\left( \frac{G_i}{\sqrt{d_A+1}}\right)+I^c_{\rho^A}\left( \frac{\mathbf{1}^A}{\sqrt{d_A+1}}\right).
$$   
The averaged correlations as the correlations relative to the quantum channel $\mathcal{E}_{\mathrm{De}}$ is defined as ~\cite{PhysRevA.105.032436}:
\begin{equation}
\begin{aligned}
Q^c_{\mathcal{E}_{\mathrm{De}}}(\rho^{AB})&:=C^c(\rho^{AB} \mid\mathcal{E}_{\mathrm{De}}\otimes \mathcal{I}_{B})-C^c(\rho^{A} \mid\mathcal{E}_{\mathrm{De}}),
\end{aligned}
\end{equation}
where $\mathcal{I}_{B}$ is the identity channel on subsystem $B$. Since $\{G_i\}_{i=1}^{d_A^2}$ is an orthonormal basis of $\mathcal{L}(\mathcal{H}_A)$, one has
$$\begin{aligned}
Q^c_{\mathcal{E}_{\mathrm{De}}}&=Q_{ob}^c(\rho^{AB}).
\end{aligned}$$

The unitary group $\mathcal U(d_A)$ on party $A$ with a $d_A$-dimensional system space $\mathcal{H}_A$ naturally induces a twirling channel,
$$
\mathcal{T}_{\mathcal U}(\rho^A)=\int_{\mathcal U(d_A)} U^\dagger \rho^A U d \mu(U),
$$
where $d \mu(U)$ is the normalized Haar measure on $\mathcal U(d_A)$. For any bipartite state $\rho^{AB}$, the corresponding quantifier of correlations in $\rho^{AB}$ relative to $\mathcal{T}_{\mathcal U}$ is defined as \cite{fan2024quantification,PhysRevA.105.032436}
\begin{align*}
Q_{\mathcal{T}_{\mathcal U}}^c(\rho^{AB})\!:=&\!\frac{d_A}{d_A+1}\int_{\mathcal U(d_A)}\! I_{\rho^{AB}}^c\left( U \otimes \mathbf{1}^B\right) \!-\!I_{\rho^A}^c\left( U\right) d \mu(U),\\
=&\! \frac{{\rm Tr}[{\tilde{c}_{f}}(L_{\rho^A},R_{\rho^A})]\!-\!\sum_{k,l}{\tilde{c}_{f}}(\lambda_k,\lambda_l){\rm{Tr}}\left[ \sigma_k^B \sigma_l^B \right]}{d_A+1}.
\end{align*}

We are now in a position to compare the averaging procedures introduced above. Therefore, whenever a complete family of mutually unbiased bases exists on $\mathcal H_A$, one has the following result.
\begin{corollary}\label{eq3.35}
For any bipartite state $\rho^{AB}$ on $\mathcal{H}_A\otimes \mathcal{H}_B$,
\begin{equation}
Q_{\rm MUB}^c(\rho^{AB})
=
Q_{\mathcal U}^c(\rho^{AB})
=
Q_{\rm ob}^c(\rho^{AB})
=
Q_{\mathcal{T}_{\mathcal U}}^c(\rho^{AB}).
\end{equation}
\end{corollary} 

Corollary~\ref{eq3.35} shows that the average correlation defined from metric-adjusted skew information is independent of the averaging scheme. Although the constructions are different in form, they give rise to the same intrinsic quantity. The MUBs formulation emphasizes complementarity, the Haar formulation emphasizes basis independence, the operator basis formulation emphasizes the observable picture, and the twirling formulation makes the underlying symmetry transparent.

Two important special cases are obtained by choosing the Wigner-Yanase skew information and the Fisher information, respectively. If the symmetric normalized operator monotone function is given by
\[
f(t)=\frac{1}{4}(1+\sqrt{t})^2,
\]
the corresponding Morozova-Chentsov function is
\[
{\tilde{c}_{f}}(x,y)=\sqrt{xy}.
\]
One has the following corollary directly according to the general expression established above.
\begin{corollary}
The average correlation associated with the Wigner-Yanase skew information has the following equivalent formulations:
\begin{equation}
Q_{\rm MUB}^{\rm WY}(\rho^{AB})
=
Q_{\mathcal U}^{\rm WY}(\rho^{AB})
=
Q_{\rm ob}^{\rm WY}(\rho^{AB})
=
Q_{\mathcal T_{\mathcal U}}^{\rm WY}(\rho^{AB}),
\label{eq3.36}
\end{equation}
and the common value is given by
\begin{equation}
Q^{\rm WY}(\rho^{AB})
=
\frac{
\bigl({\rm Tr}\sqrt{\rho^A}\bigr)^2
-
{\rm Tr}_B({\rm Tr}_A\sqrt{\rho^{AB}})^2
}{d_A+1}.
\label{eq3.37}
\end{equation}
\end{corollary}

If the metric-adjusted skew information is induced to the quantum Fisher information, that is, the symmetric normalized operator monotone function is given by
\[
f(t)=\frac{1+t}{2},
\]
the corresponding Morozova-Chentsov function is
\[
{\tilde{c}_{f}}(x,y)=\frac{2xy}{x+y}.
\]

The corresponding average correlation is given by the following result.
\begin{corollary}
The average correlation via Fisher information is given by
\begin{equation}
Q_{\rm MUB}^{F}(\rho^{AB})
=
Q_{\mathcal U}^{F}(\rho^{AB})
=
Q_{\rm ob}^{F}(\rho^{AB})
=
Q_{\mathcal T_{\mathcal U}}^{F}(\rho^{AB}),
\label{eq3.38}
\end{equation}
and the common value is
\begin{equation}
Q^{F}(\rho^{AB})
=
\frac{
\displaystyle
\sum_{i,j}\frac{2p_i p_j}{p_i+p_j}
-
\displaystyle
\sum_{k,l}\frac{2\lambda_k\lambda_l}{\lambda_k+\lambda_l}\,
{\rm{Tr}}(\sigma_k^B \sigma_l^B)}{d_A+1},
\label{eq3.39}
\end{equation}
where $p_i$ are the eigenvalues of $\rho^A$ and $\lambda_k$ are the eigenvalues of $\rho^{AB}$ and $\sigma_k^B={\rm Tr}_A(|\Psi_k\rangle\langle\Psi_k|)$ with $|\Psi_k\rangle\langle\Psi_k|$ the eigenvectors of $\rho^{AB}$.
\end{corollary}

\section{Wave-particle duality via the metric-adjusted skew information}\label{wpd}

Let $\mathcal{H}$ be a $d$-dimensional Hilbert space.The metric-adjusted skew information can be rewritten as \cite{Sun2022QuantifyingCoherence}
\begin{equation}
    \begin{aligned}I_\rho^c(A)
&=
\frac{f(0)}{2}\sum_{k,l}\frac{(p_k-p_l)^2}{p_l f(p_k/p_l)}|\langle \psi_k|A|\psi_l\rangle|^2,
    \end{aligned}
\end{equation}  
where $\rho=\sum_k p_k |\psi_k\rangle\langle\psi_k|$ is the spectral decomposition of $\rho$.

A rank one von Neumann measurement $\Pi=\{|i\rangle\langle i|: i=1,2,\ldots,d\}$ associated with an orthonormal basis $\{|i\rangle\}_{i=1}^d$  denotes the interference paths.
The wave feature can be quantified by the coherence of $\rho$ relative to $\Pi$ \cite{PhysRevA.111.052451}
\begin{equation}
    \begin{aligned}
W^c(\rho)&:=
\sum_{i=1}^d I_\rho^c(|i\rangle\langle i|)\\
&=\frac{f(0)}{2}\sum_{i=1}^d\sum_{k,l}\frac{(p_k-p_l)^2}{p_l f(p_k/p_l)}|\langle \psi_k|i\ra\la i|\psi_l\rangle|^2.
\end{aligned}
\end{equation}

The quantity $W^c(\rho)$ satisfies the following properties:

(1a) $W^c(\rho) \geqslant 0$, where the equality holds if and only if $\rho$ is a diagonal state with respect to $\Pi$.

(2a) $W^c(\rho) \leqslant 1-1 / d$, and its maximum value can be achieved if and only $\rho$ is a pure state with equal diagonal elements.

(3a) $W^c(\rho)$ is invariant under the permutation of the state index.

(4a) $W^c(\rho)$ is a convex function with respect to $\rho$.

We now introduce the particle feature relative to an orthonormal base $\{E_{ij}={|i\rangle\la j|}:i,j=1,2,\ldots,d\}$. The quantity $P^c(\rho)$ is defined as 
\begin{equation}
    \begin{aligned}P^c(\rho)
        :&=\sum_{i\neq j}^d I_\rho^c(E_{ij})\\
        &=\frac{f(0)}{2}\sum_{i\neq j}^d\sum_{k,l}\frac{(p_k-p_l)^2}{\lambda_l f(p_k/p_l)}|\langle \psi_k|E_{ij}|\psi_l\rangle|^2\\
        &=\frac{f(0)}{2}\sum_{i\neq j}^d\sum_{k,l}\frac{(p_k-p_l)^2}{\lambda_l f(p_k/p_l)}|\langle \psi_k|i\ra\la j|\psi_l\rangle|^2
    \end{aligned}
\end{equation}
The quantity $P^c(\rho)$ satisfies the following properties:

(1b) $P^c(\rho) \geqslant 0$, where the equality holds if and only if $\rho$ is the maximally mixed state.

(2b) $P^c(\rho)$ reaches its maximum value $d-1$ if and only if $\rho_{i i}=1$ for one $i$, i.e., the path is completely certain.

(3b) $P^c(\rho)$ is invariant under the permutation of the state index.

(4b) $P^c(\rho)$ is a convex function with respect to $\rho$.

The proofs of these properties are given in Appendix~\ref{app:wc} and~\ref{app:pc}, respectively. 

\section{Complementarity relations}\label{sec:complementarity}
In the context of wave-particle duality, it is natural to ask whether the wave feature and the particle feature are complementary to each other. Directly from the definitions, one has
\begin{equation}
W^c(\rho)+P^c(\rho)=\sum_{i,j}^d I_\rho^c(E_{ij})=d-{\rm Tr}[{\tilde{c}_{f}}(L_\rho,R_\rho)].
\end{equation}
According to the facts that ${\rm Tr}[{\tilde{c}_{f}}(L_\rho,R_\rho)]\geqslant 1$ and the equality holds if and only if $\rho$ is a pure state \cite{fan2023average}, we have the following result. 
\begin{proposition}\label{prop:complementarity}
For any state $\rho$ on a $d$-dimensional Hilbert space, the following complementarity relation holds:
\begin{equation}
W^c(\rho)+P^c(\rho)\leqslant d-1.
\end{equation}
The equality holds if and only if $\rho$ is a pure state.
\end{proposition}

In Ref \cite{fan2023average}, quantum $f$ entropy was introduced as
\begin{equation}
S_f(\rho)={\tilde{c}_{f}}(L_{\rho},R_{\rho})-1.
\end{equation}

Motivated by the complementarity relation via Fisher information \cite{wdxh-nwsw}, we can directly generalize it to metric-adjusted skew information and establish the complementarity relation by simple calculation.
\begin{proposition}\label{prop:complementarity2}
For any state $\rho$ on a $d$-dimensional Hilbert space, the following complementarity relation holds:
\begin{equation}
W^c(\rho)+P^c(\rho)+S_f(\rho)=d-1.
\end{equation}
\end{proposition}

Because of the function ${\tilde{c}_{f}}(x,y)$ is defined by the symmetric normalized operator monotone function $f$, our Proposition \ref{prop:complementarity2} provides a series of complementary relations for different functions of $f$. The relation shows that the wave feature, the particle feature, and the quantum $f$ entropy are complementary to each other, and their sum is a constant determined by the dimension of the Hilbert space.

We next consider the complementarity relation in the bipartite setting. Let $\rho^{AE}=|\Phi\ra\la \Phi|$ be a coupled state on quantum system $\mathcal{H}_A$ and the environment $\mathcal{H}_E$. Denote $\rho^A={\rm Tr}_E(\rho^{AE})$ and $\rho^E={\rm Tr}_A(\rho^{AE})$ the reduced density matrices on $\mathcal{H}_A$ and $\mathcal{H}_E$, respectively. In this case, the average correlation of $\rho^{AE}$ is given by
\begin{equation}
Q^c(\rho^{AE}):=\frac{{\rm Tr}[{\tilde{c}_{f}}(L_{\rho^A},R_{\rho^A})]\!-{\rm Tr}(\rho^E)^2}{d_A+1}.
\end{equation}
According to the fact 
\begin{align*}W^c(\rho^A)+P^c(\rho^A)=d_A-{\tilde{c}_{f}}(L_{\rho^A},R_{\rho^A}),\end{align*}
one derives the following complementarity relation immediately.
\begin{proposition}\label{prop:complementarity3}
For any pure state $\rho^{AE}$ on $\mathcal{H}_A\otimes \mathcal{H}_E$, the following complementarity relation holds:
\begin{equation}
W^c(\rho^A)+P^c(\rho^A)+(d_A+1)Q^c(\rho^{AE})=d_A-{\rm Tr}(\rho^E)^2.
\end{equation}
\end{proposition}

The complementarity relation in Proposition~\ref{prop:complementarity3} shows that the wave feature, the particle feature, and the average correlation are complementary to each other in the sense that their weighted sum is a constant determined by the dimension of the system and the purity of the environment. When the environment is pure, the average correlation vanishes, and the complementarity relation reduces to the standard complementarity relation for single-party system in Proposition~\ref{prop:complementarity}. This relation provides a unified description of wave-particle duality and quantum correlations in terms of metric-adjusted skew information.

\section{Conclusion}\label{sec4}

In this work, we have introduced quantum average correlation based on metric-adjusted skew information. We have examined the correlation through several natural averaging procedures, including complete families of mutually unbiased bases, all orthonormal bases and operator orthonormal bases. Our results have proved that these different approaches all lead to the same closed expression. Our work has shown that the average correlation defined from metric-adjusted skew information is an intrinsic quantity associated with the state and the chosen metric-adjusted skew information, and it is independent of the particular averaging scheme. Moreover, we have proposed wave-particle duality relations in terms of metric-adjusted skew information, and established complementarity relations between wave feature, particle feature, quantum entropy and average correlation. Therefore, our complementarity relations have provided a unified description of wave-particle duality and quantum correlations in terms of metric-adjusted skew information.

Future work may explore the operational significance of the average correlation introduced in this work, and its applications in quantum information processing tasks. It would also be interesting to investigate the behavior of the average correlation under various quantum operations, and to investigate its relationship with wave-particle duality. The extension of the present framework to multipartite systems and continuous-variable systems may also be worth exploring.

\bigskip
\noindent{\bf Acknowledgments}\, \,
This work is supported by the National Natural Science Foundation of China (NSFC) (Grant Nos.~12526648, 12401397), Natural Science Foundation of Hunan Province (Grant Nos.~2025JJ60025, 2026JJ60117), Scientific Research Project of the Education Department of Hunan Province (Grant No.~24B0298).

\bigskip
\noindent{\bf Data Availability Declarations}\, \,
No datasets were generated or analyzed during the current study.

\bigskip
\noindent{\bf Conflict of interest}\, \,
The authors declare no Conflict of interest.

    \appendix
\begin{widetext}
    \section{Proof of Proposition \ref{prop:mub}}\label{app:mub}
Let $\{|b_{t i}\rangle\}_{i=1}^{d_A}$ be the $t$-th mutually unbiased basis of $\mathcal{H}_A$. Then one has \cite{fan2023average}
$$
\sum_{t=1}^{d_A+1}\sum_{i=1}^{d_A}| b_{t i}\rangle \langle b_{t i}|\otimes |b_{t i}\rangle \langle b_{t i}|=\mathbf{1}^{A}\otimes \mathbf{1}^{A}+F,
$$
where $F$ is the swap operator on $\mathcal{H}_A\otimes \mathcal{H}_A$ defined by $F|\psi\rangle\otimes |\phi\rangle=|\phi\rangle\otimes |\psi\rangle$ for all $|\psi\rangle,|\phi\rangle\in \mathcal{H}_A$. Moreover, one has the average local coherence
\begin{align*}
&\frac{1}{d_A+1}\sum_{t=1}^{d_A+1}\sum_{i=1}^{d_A} I_{\rho^{AB}}^c\left(| b_{t i}\rangle \langle b_{t i}|\otimes \mathbf{1}^{B}\right)\\
&=\frac{1}{d_A+1}\sum_{t=1}^{d_A+1}\sum_{i=1}^{d_A} {\rm Tr} \left(\rho^{AB} | b_{t i}\rangle \langle b_{t i}|\otimes \mathbf{1}^{B} \right)-\frac{1}{d_A+1}\sum_{t=1}^{d_A+1}\sum_{i=1}^{d_A} {\rm Tr}\left((| b_{t i}\rangle \langle b_{t i}| \otimes \mathbf{1}^{B})  {\tilde{c}_{f}}(L_{\rho^{AB}},R_{\rho^{AB}})(| b_{t i}\rangle \langle b_{t i}| \otimes \mathbf{1}^{B}) \right )\\
&=1-\frac{1}{d_A+1}\sum_{t=1}^{d_A+1}\sum_{i=1}^{d_A}\sum_{k,l}{\tilde{c}_{f}}(\lambda_k,\lambda_l){\rm Tr}\left( (| b_{t i}\rangle \langle b_{t i}|\otimes \mathbf{1}^{B}) |\Psi_k\rangle\langle\Psi_k| (| b_{t i}\rangle \langle b_{t i}| \otimes \mathbf{1}^{B}) |\Psi_l\rangle\langle\Psi_l|\right)\\
&=1-\frac{1}{d_A+1}\sum_{t=1}^{d_A+1}\sum_{i=1}^{d_A}\sum_{k,l}\sum_{\mu ,\nu}\sum_{\mu^\prime ,\nu^\prime}{\tilde{c}_{f}}(\lambda_k,\lambda_l){\rm Tr}\left( (| b_{t i}\rangle \langle b_{t i}|\otimes \mathbf{1}^{B} )(X_{\mu \nu}^k \otimes |\mu^k\ra \la \nu^k|) (| b_{t i}\rangle \langle b_{t i}| \otimes \mathbf{1}^{B} )(X_{\mu^\prime \nu^\prime}^l \otimes |{\mu^\prime}^l\ra \la {\nu^\prime}^l|)\right)\\
&=1-\frac{1}{d_A+1}\sum_{k,l}\sum_{\mu ,\nu}\sum_{\mu^\prime ,\nu^\prime}{\tilde{c}_{f}}(\lambda_k,\lambda_l)\sum_{t=1}^{d_A+1}\sum_{i=1}^{d_A}{\rm Tr}\left( | b_{t i}\rangle \langle b_{t i}| X_{\mu \nu}^k | b_{t i}\rangle \langle b_{t i}| X_{\mu^\prime \nu^\prime}^l\right)   {\rm Tr}\left(|\mu^k\ra \la \nu^k|(|{\mu^\prime}^l\ra \la {\nu^\prime}^l|)\right)\\
&=1-\frac{1}{d_A+1}\sum_{k,l}\sum_{\mu ,\nu}\sum_{\mu^\prime ,\nu^\prime}{\tilde{c}_{f}}(\lambda_k,\lambda_l){\rm Tr}\left(|\mu^k\ra \la \nu^k|(|{\mu^\prime}^l\ra \la {\nu^\prime}^l|)\right)\sum_{t=1}^{d_A+1}\sum_{i=1}^{d_A}{\rm Tr}\left( (| b_{t i}\rangle \langle b_{t i}|\otimes |b_{t i}\rangle \langle b_{t i}|) (X_{\mu \nu}^k \otimes X_{\mu^\prime \nu^\prime}^l)\right)\\
&=1-\frac{1}{d_A+1}\sum_{k,l}\sum_{\mu ,\nu}\sum_{\mu^\prime ,\nu^\prime}{\tilde{c}_{f}}(\lambda_k,\lambda_l){\rm Tr}\left(|\mu^k\ra \la \nu^k|(|{\mu^\prime}^l\ra \la {\nu^\prime}^l|)\right){\rm Tr}\left((\mathbf{1}^{A}\otimes \mathbf{1}^{A}+F) (X_{\mu \nu}^k \otimes X_{\mu^\prime \nu^\prime}^l)\right)\\
&=1-\frac{1}{d_A+1}\sum_{k,l}\sum_{\mu ,\nu}\sum_{\mu^\prime ,\nu^\prime}{\tilde{c}_{f}}(\lambda_k,\lambda_l){\rm Tr}\left(|\mu^k\ra \la \nu^k|(|{\mu^\prime}^l\ra \la {\nu^\prime}^l|)\right)\left({\rm Tr}\left(X_{\mu \nu}^k\right){\rm Tr}\left(X_{\mu^\prime \nu^\prime}^l\right)+{\rm Tr}\left( X_{\mu \nu}^k X_{\mu^\prime \nu^\prime}^l\right)\right)\\
&=1-\frac{1}{d_A+1}\sum_{k,l}{\tilde{c}_{f}}(\lambda_k,\lambda_l)\left({\rm Tr}\left({\rm{Tr}}_A\left(|\Psi_k\rangle\langle\Psi_k|\right){\rm{Tr}}_A\left(|\Psi_l\rangle\langle\Psi_l|\right)\right)+|\langle\Psi_k|\Psi_l\rangle|^2\right)\\
&=1-\frac{1}{d_A+1}\sum_{k,l}{\tilde{c}_{f}}(\lambda_k,\lambda_l){\rm Tr}\left({\rm{Tr}}_A\left(|\Psi_k\rangle\langle\Psi_k|\right){\rm{Tr}}_A\left(|\Psi_l\rangle\langle\Psi_l|\right)\right)-\frac{1}{d_A+1}\sum_{k}{\tilde{c}_{f}}(\lambda_k,\lambda_k)\\
&=\frac{d_A-\sum_{k,l}{\tilde{c}_{f}}(\lambda_k,\lambda_l){\rm Tr}\left({\rm{Tr}}_A\left(|\Psi_k\rangle\langle\Psi_k|\right){\rm{Tr}}_A\left(|\Psi_l\rangle\langle\Psi_l|\right)\right)}{d_A+1}\\
&=\frac{d_A-\sum_{k,l}{\tilde{c}_{f}}(\lambda_k,\lambda_l){\rm Tr}\left( \sigma_k^B \sigma_l^B \right)}{d_A+1}.
\end{align*}

Combining Eqs.~\eqref{eq2.15}, we complete the proof of Proposition \ref{prop:mub}. $\square$

\section{Proof of Proposition \ref{prop:haar}}\label{app:haar}
Following the relation \cite{zhang2024matrixintegralsunitarygroups}
\begin{equation*}\label{zhanglinunitary}
\begin{aligned} 
\int_{\mathcal{U}(d)} {U}^{\dagger} {A} {U} {X} {U}^{\dagger} {B} {U} d\mu(U) =& \frac{d \mathbf{tr}({A} {B})-\mathbf{tr}({A}) \mathbf{tr}({B})}{d\left(d^2-1\right)} \mathbf{tr}({X}) \mathbf{1}+\frac{d \mathbf{tr}({A}) \mathbf{tr}({B})-\mathbf{tr}({A} {B})}{d\left(d^2-1\right)} {X},
\end{aligned}
\end{equation*}
where $A$, $B$ and $X$ are operators with $d$ dimension, $\mathbf{1}$ is the identity matrix. we evaluate the average local coherence based on all orthogonal bases
\begin{align*}
&\int_{\mathcal U(d_A)}\sum_i I(\rho^{AB},U|i^A\rangle\langle i^A|U^\dagger\otimes \mathbf{1}^{B})d\mu(U)\\
&=\sum_i \int_{\mathcal U(d_A)}{\rm{Tr}}\left(\rho^{AB}(U|i^A\rangle\langle i^A|U^\dagger\otimes \mathbf{1}^{B})\right)-{\rm{Tr}}\left((U|i^A\rangle\langle i^A|U^\dagger\otimes \mathbf{1}^{B})  {\tilde{c}_{f}}(L_{\rho^{AB}},R_{\rho^{AB}}) (U|i^A\rangle\langle i^A|U^\dagger\otimes \mathbf{1}^{B})\right)d\mu(U)\\
&=1-\sum_i\sum_{k,l}{\tilde{c}_{f}}(\lambda_k,\lambda_l)\int_{\mathcal U(d_A)}{\rm{Tr}}\left((U|i^A\rangle\langle i^A|U^\dagger\otimes \mathbf{1}^{B}) |\Psi_k\rangle\langle\Psi_k|  (U|i^A\rangle\langle i^A|U^\dagger\otimes \mathbf{1}^{B})|\Psi_l\rangle\langle\Psi_l|\right)d\mu(U)\\
&=1-\sum_i\sum_{k,l}{\tilde{c}_{f}}(\lambda_k,\lambda_l)\\
&\quad  \times\sum_{\mu ,\nu}\sum_{\mu^\prime ,\nu^\prime}\int_{\mathcal U(d_A)}\!\!\!\!\!\!{\rm{Tr}}\left((U|i^A\rangle\langle i^A|U^\dagger\otimes \mathbf{1}^{B}) (X_{\mu \nu}^k \otimes |\mu^k\ra \la \nu^k|)  (U|i^A\rangle\langle i^A|U^\dagger\otimes \mathbf{1}^{B})(X_{\mu^\prime \nu^\prime}^l \otimes |{\mu^\prime}^l\ra \la {\nu^\prime}^l|)\right)d\mu(U)\\
&=1-\sum_i\sum_{k,l}{\tilde{c}_{f}}(\lambda_k,\lambda_l)\sum_{\mu ,\nu}\sum_{\mu^\prime ,\nu^\prime}\int_{\mathcal U(d_A)}{\rm{Tr}}(U|i^A\rangle\langle i^A|U^\dagger X_{\mu \nu}^k U|i^A\rangle\langle i^A|U^\dagger X_{\mu^\prime \nu^\prime}^l)d\mu(U) {\rm Tr}\left(|\mu^k\ra \la \nu^k|(|{\mu^\prime}^l\ra \la {\nu^\prime}^l|)\right)\\
&=1-\sum_{k,l}{\tilde{c}_{f}}(\lambda_k,\lambda_l)\sum_{\mu ,\nu}\sum_{\mu^\prime ,\nu^\prime} {\rm Tr}\left(|\mu^k\ra \la \nu^k|(|{\mu^\prime}^l\ra \la {\nu^\prime}^l|)\right) \sum_i{\rm{Tr}}\left (|i^A\rangle\langle i^A| \int_{\mathcal U(d_A)}U^\dagger X_{\mu \nu}^k U|i^A\rangle\langle i^A|U^\dagger X_{\mu^\prime \nu^\prime}^l Ud\mu(U)\right) \\
&=1-\sum_{k,l}{\tilde{c}_{f}}(\lambda_k,\lambda_l)\sum_{\mu ,\nu}\sum_{\mu^\prime ,\nu^\prime} {\rm Tr}\left(|\mu^k\ra \la \nu^k|(|{\mu^\prime}^l\ra \la {\nu^\prime}^l|)\right)\sum_i {\rm{Tr}}\left (|i^A\rangle\langle i^A|\frac{d_A{\rm{Tr}}(X_{\mu \nu}^k X_{\mu^\prime \nu^\prime}^l)-{\rm{Tr}}(X_{\mu \nu}^k){\rm{Tr}}( X_{\mu^\prime \nu^\prime}^l)}{d_A (d_A^2-1)}\right)\\
&\quad-\sum_{k,l}{\tilde{c}_{f}}(\lambda_k,\lambda_l)\sum_{\mu ,\nu}\sum_{\mu^\prime ,\nu^\prime} {\rm Tr}\left(|\mu^k\ra \la \nu^k|(|{\mu^\prime}^l\ra \la {\nu^\prime}^l|)\right)\sum_i {\rm{Tr}}\left (|i^A\rangle\langle i^A|\frac{d_A{\rm{Tr}}(X_{\mu \nu}^k){\rm{Tr}}( X_{\mu^\prime \nu^\prime}^l)-{\rm{Tr}}(X_{\mu \nu}^k X_{\mu^\prime \nu^\prime}^l)}{d_A (d_A^2-1)}\right )\\
&=1-\sum_{k,l}{\tilde{c}_{f}}(\lambda_k,\lambda_l)\sum_{\mu ,\nu}\sum_{\mu^\prime ,\nu^\prime} {\rm Tr}\left(|\mu^k\ra \la \nu^k|(|{\mu^\prime}^l\ra \la {\nu^\prime}^l|)\right)\frac{{\rm{Tr}}(X_{\mu \nu}^k){\rm{Tr}}( X_{\mu^\prime \nu^\prime}^l)+{\rm{Tr}}(X_{\mu \nu}^k X_{\mu^\prime \nu^\prime}^l)}{d_A+1}\\
&=\frac{d_A-\sum_{k,l}{\tilde{c}_{f}}(\lambda_k,\lambda_l){\rm Tr}\left({\rm{Tr}}_A\left(|\Psi_k\rangle\langle\Psi_k|\right){\rm{Tr}}_A\left(|\Psi_l\rangle\langle\Psi_l|\right)\right)}{d_A+1}\\
&=\frac{d_A-\sum_{k,l}{\tilde{c}_{f}}(\lambda_k,\lambda_l){\rm Tr}\left( \sigma_k^B \sigma_l^B \right)}{d_A+1}.
\end{align*}
Complete the proof by combining Eq.~\eqref{eq2.15}. $\square$

\section{Proof of properties of $W^c(\rho)$} \label{app:wc} 
(1a) $W^c(\rho) \geqslant 0$ follows from the non-negativity of the metric-adjusted skew information. If $W^c(\rho)=0$, then $I_\rho^c(|i\rangle\langle i|)=0$ for all $i$. This implies that $\rho$ commutes with each $|i\rangle\langle i|$, which means $\rho$ is diagonal in the basis defined by $\Pi$. Conversely, if $\rho$ is diagonal in the basis defined by $\Pi$, then $I_\rho^c(|i\rangle\langle i|)=0$ for all $i$, which implies $W^c(\rho)=0$.

(2a) According to the fact that the metric-adjusted skew information is convex, the maximum value of $W^c(\rho)$ is achieved when $\rho$ is a pure state $|\psi\ra$, that is
\begin{align*}
W^c(|\psi\ra\la \psi|) &= 1 - \sum_i|\la i|\psi\ra\la \psi | i\ra|^2\leqslant 1 - \frac{1}{d}.
\end{align*}
 The equality holds if and only if $|\psi\ra$ has equal diagonal elements, which corresponds to a maximally coherent state.

(3a) The invariance under permutation of the state index follows from the definition of $W^c(\rho)$, which depends only on the eigenvalues and eigenvectors of $\rho$ and not on their ordering.

(4a) The convexity of $W^c(\rho)$ can be shown by using the convexity of the metric-adjusted skew information. For any two states $\rho_1$ and $\rho_2$, and any $0 \leqslant \lambda \leqslant 1$, we have
\begin{align*}
W^c(\lambda \rho_1 + (1-\lambda) \rho_2) &= \sum_i I_{\lambda \rho_1 + (1-\lambda) \rho_2}^c(|i\rangle\langle i|) \\
&\leqslant \lambda \sum_i I_{\rho_1}^c(|i\rangle\langle i|) + (1-\lambda) \sum_i I_{\rho_2}^c(|i\rangle\langle i|) \\
&= \lambda W^c(\rho_1) + (1-\lambda) W^c(\rho_2).
\end{align*}
This completes the proof of the properties of $W^c(\rho)$. $\square$

\section{Proof of properties of $P^c(\rho)$} \label{app:pc}

(1b) $P^c(\rho) \geqslant 0$ follows from the non-negativity of the metric-adjusted skew information. If $P^c(\rho)=0$, then $I_\rho^c(E_{ij})=0$ for all $i\neq j$. This implies that $\rho$ commutes with each $E_{ij}$, which means $\rho$ is diagonal in the basis defined by $\{E_{ij}\}$, i.e., $\rho$ is the maximally mixed state. Conversely, if $\rho$ is the maximally mixed state, then $I_\rho^c(E_{ij})=0$ for all $i\neq j$, which implies $P^c(\rho)=0$.

(2b) According to the fact that the metric-adjusted skew information is convex, the maximum value of $P^c(\rho)$ is achieved when $\rho$ is a pure state $|\psi\ra$, that is
\begin{align*}P^c(|\psi\ra\la \psi|) &= d-1 - \sum_{i\neq j}|\la i|\psi\ra\la \psi | j\ra|^2\leqslant d-1.
\end{align*}    
The maximum value of $P^c(\rho)$ is achieved if and only if $\rho$ is a pure state with one diagonal element equal to 1 and the rest equal to 0, which corresponds to a state with complete path certainty.

(3b) The invariance under permutation of the state index follows from the definition of $P^c(\rho)$, which depends only on the eigenvalues and eigenvectors of $\rho$ and not on their ordering.

(4b) The convexity of $P^c(\rho)$ can be shown by using the convexity of the metric-adjusted skew information. For any two states $\rho_1$ and $\rho_2$, and any $0 \leqslant \lambda \leqslant 1$, we have
\begin{align*}P^c(\lambda \rho_1 + (1-\lambda) \rho_2) &= \sum_{i\neq j} I_{\lambda \rho_1 + (1-\lambda) \rho_2}^c(E_{ij}) \\
&\leqslant \lambda \sum_{i\neq j} I_{\rho_1}^c(E_{ij}) + (1-\lambda) \sum_{i\neq j} I_{\rho_2}^c(E_{ij}) \\
&= \lambda P^c(\rho_1) + (1-\lambda) P^c(\rho_2). 
\end{align*}This completes the proof of the properties of $P^c(\rho)$. $\square$

\end{widetext}

\bibliography{zhang}
\bibliographystyle{iopart-num}

\end{document}